\documentclass[reprint,pre,showpacs,amsmath,amssymb,aps]{revtex4-1}
\usepackage{natbib}
\usepackage{graphicx,color,rotating}
\usepackage[latin1]{inputenc}   %Hvorfor bliver jeg nødt til at udkommentere denne?
\usepackage{textcomp}
\usepackage{dcolumn}% Align table columns on decimal point
\usepackage{bm}     % bold math
\usepackage[version=3]{mhchem}  %For chemistry

\newcommand{\notop}{{{}_{}}}

\newcommand{\mr}[1]{\ensuremath{\mathrm{#1}}}

\newcommand{\pp}{\partial}       % This used to be \partial^{{}}

\newcommand{\JH}{J_{\mathrm{H}}}
\newcommand{\JOH}{J_{\mathrm{OH}}}
\newcommand{\Jw}{J_{\mathrm{w}}}

\newcommand{\ca}{c^{{}}_\mathrm{a}}
\newcommand{\cacrit}{c^{\mathrm{crit}}_\mathrm{a}}

\newcommand{\cH}{c^{\mathstrut}_{\mathrm{H}}}

\newcommand{\cOH}{c^{\mathstrut}_{\mathrm{OH}}}

\newcommand{\cw}{c_{\mathrm{w}}}
\newcommand{\cb}{c_{\mathrm{b}}}

\newcommand{\kB}{{k^\notop_\mathrm{B}}}
\newcommand{\kBT}{{k^\notop_\mathrm{B} T}}

\newcommand{\Nm}{N_{\mathrm{m}}}

\newcommand{\lamD}{\lambda^{{}}_\mathrm{D}}
\newcommand{\blamD}{\bar{\lambda}^{{}}_\mathrm{D}}

\newcommand{\blamDsqr}{\bar{\lambda}^{{2}}_\mathrm{D}}
\newcommand{\blamDinv}{\bar{\lambda}^{-1_{}}_\mathrm{D}}

\newcommand{\rhoe}{\rho^\notop_\mathrm{el}}

\newcommand{\VT}{V_{\mathrm{T}}}
\newcommand{\Eh}{\hat{E}}
\newcommand{\phih}{\hat{\phi}}
\newcommand{\rhoh}{\hat{\rho}_{\mathrm{el}}}
\newcommand{\xh}{\hat{x}}
\newcommand{\VO}{V^{{}}_0}

\newcommand{\xrho}{x_{\rho}}
\newcommand{\SIC}{\textrm{C}}

\newcommand{\SIF}{\textrm{F}}

\newcommand{\SIm}{\textrm{m}}
\newcommand{\SIM}{\textrm{M}}
\newcommand{\SImM}{\textrm{mM}}

\newcommand{\SImm}{\textrm{mm}}
\newcommand{\SImum}{\textrm{\textmu{}m}}

\newcommand{\SImV}{\textrm{mV}}

\newcommand{\beq}[1]{\begin{equation} \eqlab{#1}}
\newcommand{\eeq}{\end{equation}}
\newcommand{\bsub}{\begin{subequations}}
\newcommand{\esub}{\end{subequations}}
\def\bal#1\eal{\begin{align}#1\end{align}}
\def\bsubal#1\esubal{\bsub \begin{align}#1\end{align} \esub}

\newcommand{\eqlab}[1]{\label{eq:#1}}
\renewcommand{\eqref}[1]{Eq.~(\ref{eq:#1})}
\newcommand{\eqsref}[2]{Eqs.~(\ref{eq:#1}) and~(\ref{eq:#2})}
\newcommand{\eqsthreeref}[3]{Eqs.~(\ref{eq:#1}), (\ref{eq:#2}), and (\ref{eq:#3})}
\newcommand{\eqsfourref}[4]{Eqs.~(\ref{eq:#1}), (\ref{eq:#2}), (\ref{eq:#3}), and (\ref{eq:#4})}

\newcommand{\figref}[1]{Fig.~\ref{fig:#1}}

\newcommand{\figlab}[1]{\label{fig:#1}}

\newcommand{\secref}[1]{Section~\ref{sec:#1}}

\newcommand{\seclab}[1]{\label{sec:#1}}
\newcommand{\tabref}[1]{Table~\ref{tab:#1}}

\newcommand{\tablab}[1]{\label{tab:#1}}

\begin{document}
%\preprint{Preprint identifier}

%\title{Transport-limited water splitting in systems exhibiting concentration polarization}	

\title{Transport-limited water splitting at ion-selective interfaces\\
 during concentration polarization}	

\author{Christoffer P. Nielsen}
\affiliation{Department of Physics, Technical University of Denmark, DTU Physics Building 309, DK-2800 Kongens Lyngby, Denmark}
\email{chnie@fysik.dtu.dk, bruus@fysik.dtu.dk}

\author{Henrik Bruus}
\affiliation{Department of Physics, Technical University of Denmark, DTU Physics Building 309, DK-2800 Kongens Lyngby, Denmark}

\date{11 December 2013}

\begin{abstract}

We present an analytical model of salt- and water-ion transport across an ion-selective interface based on an assumption of local equilibrium of the water-dissociation reaction. The model yields current-voltage characteristics and curves of water-ion current versus salt-ion current, which are in qualitative agreement with experimental results published in the literature. The analytical results are furthermore in agreement with direct numerical simulations. As part of the analysis, we find approximate solutions to the classical problem of pure salt transport across an ion-selective interface. These solutions provide closed-form expressions for the current-voltage characteristics, which include the overlimiting current due to the development of an extended space charge region. Finally, we discuss how the addition of an acid or a base affects the transport properties of the system and thus provide predictions accessible to further experimental tests of the model.
\end{abstract}

%\keywords{Water splitting \and Concentration polarization \and Electrokinetics \and Ion-selective interface \and Ion-selective membrane \and Ion-exchange membrane \and Electrodialysis}
% 47.57.jd 	Electrokinetic effects
% 82.39.Wj 	Ion exchange, dialysis, osmosis, electro-osmosis, membrane processes
%82.45.Mp Thin layers, films, monolayers, membranes in electrochemistry
%66.10.-x Diffusion and ionic conduction in liquids

\maketitle

\section{Introduction}
\seclab{Intro}
Ion transport across an ion-selective interface, such as a nanochannel, an electrode or an ion-selective membrane, has found numerous applications in e.g.\ dialysis, desalination, battery and fuel cell technology, electrochemistry, and microfluidic systems \cite{Nikonenko2010, Kim2010a, Park2010, Winter2004, Malek2013, Tanaka2012, Etacheri2011}. A common feature of ion transport across ion-selective interfaces is the phenomenon known as concentration polarization, in which the ion-concentration undergoes depletion next to the interface leading to a decrease in conductivity \cite{Nikonenko2010}. In the classical one-dimensional local electro-neutrality (LEN) modeling of the problem, the conductivity goes to zero as the voltage drop over the system is increased, and the current approaches the so-called limiting current. Experimentally it has however been found that many concentration-polarized systems can sustain a significant overlimiting current \cite{Nikonenko2010, Taky1992, Maletzki1992}. A number of mechanisms have been suggested as explanation for this overlimiting current: these include the development of an extended space charge region (ESC) \cite{Smyrl1967,Rubinstein1979, Yariv2009}, electroosmotic instabilities (EOI) \cite{Rubinstein2000, Rubinstein2002}, water splitting \cite{Rubinstein1977, Kharkats1979}, current-induced membrane discharge (CIMD) \cite{Andersen2012}, and surface conduction in microchannels \cite{Dydek2011}. Increasing amounts of evidence points to EOI as the primary mechanism in systems where advection is not suppressed by the geometry \cite{Rubinstein2002, Maletzki1992}. However, because of the experimental and theoretical difficulties associated with investigating concentration polarization, no unified picture describing the relative importance of mechanisms in different regimes has yet emerged. Concentration polarization is therefore still very much an open problem, warranting additional investigations into the underlying mechanisms.

In this paper we investigate the effect of water splitting and an extended space-charge region on systems exhibiting concentration polarization. Apart from being relevant for classical concentration polarization in macroscopic systems, our investigation of water splitting is motivated by the recent studies which highlight the importance of reactions between hydronium and surface groups in microsystems \cite{Behrens2001,Lint2002,Jensen2011,Andersen2011}.

%The classical one-dimensional local electro-neutrality (1D LEN) modeling of the problem has proven to fall short in many ways, since neither the effect of a finite space charge nor the influence of water ions (hydronium and hydroxide) is accounted for in this model \cite{Nikonenko2010, Smyrl1967, Chu2005}.

\begin{table}[!t]
\caption{\tablab{abbrev} List of abbreviations used in this work.}
\centering
\begin{ruledtabular}
\begin{tabular}{l c}
Concept & Abbreviation \\
\hline
Local electro-neutrality & LEN \\
Space-charge region & SCR \\
Extended space-charge region & ESC \\
Electric double layer & EDL \\
Electroosmotic instability & EOI \\
% Electric double layer & EDL\\
\end{tabular}
\end{ruledtabular}
\end{table}

\begin{figure*}[!t]
    \includegraphics[]{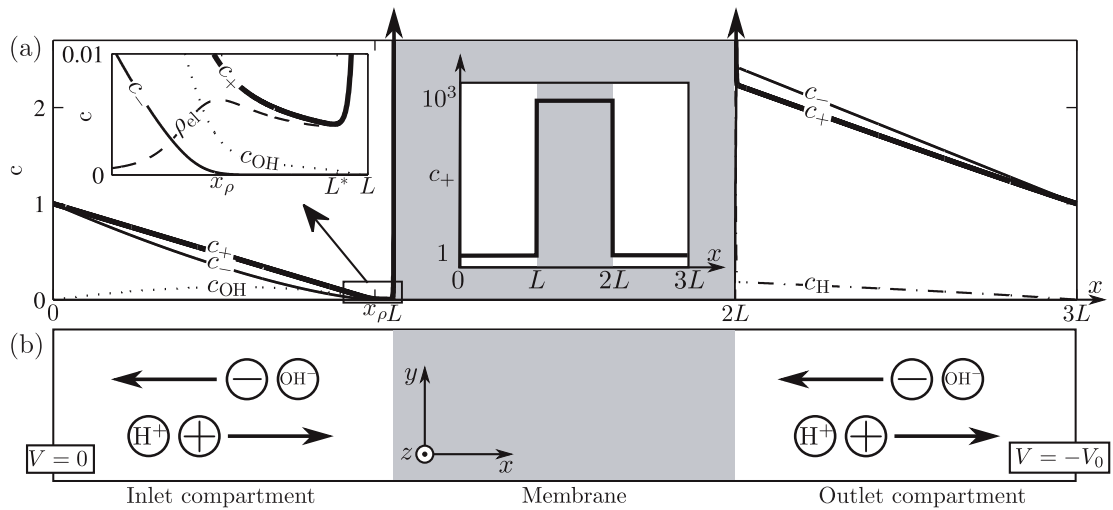}
    \caption{\figlab{System2} (a) Normalized concentrations of salt ions ($c_+$ and $c_-$) and water ions ($c_{\mr{OH}}$ and $c_{\mr{H}}$) obtained from a numerical simulation, see \secref{numerics}. The top left insert is a zoom of the space-charge region near $x_{\rho}$ in front of the membrane. The center insert is a plot of the normalized cation concentration $c_+$ showing the very high concentration inside the membrane $L<x<2L$ (gray). (b) Sketch of the studied system with salt ions ($+$ and $-$) and water ions ($\mr{H}^+$ and $\mr{OH}^-$). An inlet compartment ($0 < x < L$) and an outlet compartment ($2L<x<3L$) separated by an ion-selective nanoporous membrane. To the left ($x=0$) and right ($x=3L$) the system is connected to reservoirs of well-defined salt concentration and pH. The system is considered to be translationally invariant in the $yz$-plane parallel to the membrane.}
\end{figure*}

Water splitting has long been investigated as a possible cause of overlimiting current in systems exhibiting concentration polarization \cite{Block1966,Sonin1972,Rubinstein1977}. In 1979 Kharkats predicted that besides adding to the total current in the system, a water-ion current is also able to increase, or exalt, the current of salt ions above the limiting current \cite{Kharkats1979}. Since then, the effect and origin of the water-ion current has drawn considerable attention, and experiments have largely confirmed the fundamentals of Kharkats prediction \cite{Simons1985, Nikonenko2005, Zabolotsky1998, Tanaka1986,Mavrov1993}. It is reasonably well understood that the origin of the water-ion current is water dissociation taking place in a region close to the ion-selective interface. In many experiments the magnitude of the water-ion current does however indicate a reaction rate much larger than what should be possible, considering only the bulk dissociation rates \cite{Nikonenko2010, Tanaka2010}. A number of models have been suggested to explain this remarkable feature. Some of these ascribe the increased reaction rate to catalytic interactions with membrane surface groups \cite{Simons1979, Simons1985,Jialin1998}, while others use that the dissociation rate is increased in strong electric fields and employ a phenomenological function with one or more fitting parameters to describe this dependence \cite{Simons1984, Danielsson2009, Nikonenko2005, Tanaka2010}. In lack of conclusive evidence in support of either theory, the only thing that can be said with some confidence is that the actual reaction kinetics are probably exceedingly complicated.

In this work we avoid the subject of the detailed reaction kinetics altogether by simply assuming that the dissociation rate is so large that the water-ion current is transport limited rather than reaction limited. Put in another way, we assume local equilibrium of the water-ions everywhere in the system as done in Refs.~\cite{Zabolotskii2002, Andersen2012}. Since the analysis given in this paper is based on this assumption, experiments supporting our conclusions would serve to corroborate the underlying assumption of local equilibrium of the water-dissociation reaction. In particular, the techniques allowing for individual measurements of salt current and water-ion current, such as titration-based methods \cite{Zabolotsky1998, Nikonenko2005}, are highly relevant, as many of our results and predictions depend explicitly on both these currents.

Even for systems where the water-equilibrium assumption is not justified, the presented analysis is valuable, since it provides an upper bound to the currents which can be obtained (assuming that the equilibrium constant $K_{\mr{w}}$ remains fixed). Also, since the developed model employs a minimum of assumptions about the system, it is an excellent model to benchmark more detailed reaction models against. It has for instance been a success criterion for reaction models that they are able to replicate the characteristic S-shape (increase-plateau-increase, see \secref{NumRes}) of the experimental current-voltage curves \cite{Danielsson2009}. However, such S-shaped current-voltage curves are found even in our simple model, which suggests that they are a result of the transport properties of the system rather than the detailed reaction kinetics.

To simplify the treatment and bring forth the fundamental physics of water splitting, we study a system which is translationally invariant parallel to the ion-selective interface, and we use a 1D model to describe this essentially one-dimensional system. By employing a 1D model we disregard the possibility of spontaneous symmetry-breaking, occurring at higher voltages in the form of electroosmotic instabilities (EOI) \cite{Rubinstein2000, Rubinstein2002}, as this effect can only be described in a full 3D model. For a number of systems where advection is suppressed by gels, microchannels or porous structures disregarding EOI is actually justified, and even when that is not the case our model provides a way to study the behavior before EOI sets in as well as the transition to EOI.
%To make matters as simple as possible we use a 1D model and do not consider the possibility of electroosmotic instabilities (EOI). Apart from being a convenient assumption for a preliminary study this assumption applies directly to a number of situations where advection is suppressed by gels, microchannels or porous structures.

As a concrete realization of an ion-selective interface we investigate an ion-selective membrane. The employed methods are however completely general, and most of the conclusions carry over to transport across any ion-selective interface.

 The most common abbreviations used in this work are listed in \tabref{abbrev}.

\section{The model system}
\seclab{model}

The one-dimensional model system stretching along the $x$-axis is shown in \figref{System2}. It consists of a central ion-selective membrane of length $L$ connected to two well-mixed reservoirs, to the left and right, through two compartments each also of length $L$. The reservoirs have well-defined salt concentration $c_0$ and pH, and there is a potential difference $V_0$ between them. The system is translationally invariant in the $yz$-plane parallel to the membrane. In \figref{System2}(a)  are shown typical concentration distributions obtained from the numerical simulations described in \secref{numerics}. The top left insert shows the ion and charge concentrations in the space-charge region (SCR) near the membrane and three points $\xrho$, $L^*$ and $L$ are defined for later use: $\xrho$ denotes the position of the peak in space charge density, $L^*$ denotes the beginning of the quasi-equilibrium electric double layer (EDL), and $L$ is the length of the left compartment. Inside the membrane, the concentration of anions vanishes while the concentration of cations becomes very large ($\sim 10^3$ times the reservoir concentration, depending on system parameters). In \figref{System2}(b) is shown a sketch of the model system. The ions in the model are positive and negative salt ions with concentration $c_+$ and $c_-$, respectively, as well as hydronium and hydroxide ions (water ions) with concentration $\cH$ and $\cOH$, respectively.

\section{Governing equations}
\seclab{Gov_eq}

We consider monovalent ions and normalize the ion concentrations by the reservoir salt concentration $c_0 = c_+(0) = c_-(0)$. The electrical potential $\phi$ is normalized by the thermal voltage $V_{\mr{T}} = \kBT/e$ and the position by the length $L$. The cation current is normalized by the classical limiting current $J_{\mr{lim}} = 2D_+c_0/L$, the anion current is normalized by $2D_-c_0/L$ while $\JH$ and $\JOH$ are both normalized by $2D_{\mr{OH}}c_0/L$. The non-dimensionalized ion-currents are
\bsub
\eqlab{Currents}
\bal
2J_+ &= -\pp_x c_+ - c_+\pp_x\phi,\eqlab{Jp} \\
2J_- &= -\pp_x c_- + c_-\pp_x\phi, \eqlab{Jm}\\
2\JH &= -\beta\pp_x \cH - \beta\cH\pp_x\phi, \\
2\JOH &= -\pp_x \cOH + \cOH\pp_x\phi,
\eal
\esub
where we have introduced the diffusivity ratio  $\beta\equiv D_{\mr{H}}/D_{\mr{OH}} = 1.75$. In the remainder of the paper we are primarily concerned with non-dimensional quantities. For the rare exceptions of dimension-full quantities, these will be indicated by a tilde.

In steady state the relevant Nernst--Planck equations for the salt ions are
\begin{equation} \eqlab{Salt_NP}
\pp_x J_{\pm} =0.
\end{equation}
Similar equations govern the motion of hydronium and hydroxide, with the addition of a reaction term $R$, which derives from the auto-protolytic reaction of water
\bsubal
0 &= -\pp_x \JH +R,\\
0 &= -\pp_x \JOH +R.
\esubal
Here the reaction rates are identical since the reaction
\begin{equation}
\ce{ H3O+ + OH- <=> 2H2O},
\end{equation}
produces or consumes one unit of each species. Introducing the water-ion current $\Jw\equiv \JH-\JOH$ we obtain a single transport equation for the water ions
 \begin{equation}  \eqlab{Water_NP}
 \pp_x \Jw = 0.
 \end{equation}
The fundamental assumption in this work is that the time scale of the auto-protolysis is much shorter than the transport time of hydronium and hydroxide. That is, we work in the limit of high Damk\"{o}hler number, for which the hydronium and hydroxide concentrations are simply related via the equilibrium constant $K_{\mr{w}} = \tilde{c}_{\ce{OH}}\tilde{c}_{\ce{H}}$, which for dimensionless concentrations can be written as
\begin{equation}   \eqlab{Water_LEQ}
\cOH\:\cH = n^2, \text{ with } n = \frac{\sqrt{K_{\mr{w}}}}{c_0}.
\end{equation}
The final governing equation is the Poisson equation
\bsubal
\eqlab{Poisson}
2\blamDsqr \pp_x^2 \phi &= -c_++c_--\cH+\cOH,\\
\blamD &\equiv \frac{\lamD}{L} = \frac{1}{L}\:\sqrt{\frac{\epsilon_{\mr{w}} V_{\mr{T}}}{2ec_0}},
\esubal
where the nondimensionalized Debye length $\blamD$ has been introduced, with  $e$ being the unit charge and $\epsilon_{\mr{w}}$ the permittivity of water. Since $\blamDsqr$ is a small parameter any small deviation from charge neutrality will greatly affect the potential in a manner which tends to restore charge neutrality. This observation is the basis of the often used local electro-neutrality (LEN) assumption, where the bulk liquid is assumed electro-neutral and the only deviation from electro-neutrality is in the Debye layer.

The membrane is modelled as having a high density $N_{\mr{m}}$ of frozen negative charges (normalized by $c_0$), a porosity $\epsilon_{\mr{P}}$, a permittivity $\epsilon_{\mr{m}}$ and a tortuosity $\tau$. Inside the membrane the currents and the Poisson equation are therefore modified as
\bsub \eqlab{Membrane_eq}
\bal
2J_i &= \frac{\epsilon_{\mr{P}}}{\tau}(-\pp_x c_i \pm c_i\pp_x \phi),\\
\frac{\epsilon_{\mr{m}}}{\epsilon_{\mr{w}}}2\blamDsqr \pp_x^2 \phi &= \epsilon_{\mr{P}}(-c_++c_--\cH+\cOH)+N_{\mr{m}}.
\eal
\esub
Most ion-selective membranes have a complex structure \cite{heitner-wirguin1996, Xu2005, Nagarale2006} making it difficult to properly determine the values of $N_{\mr{m}}$, $\epsilon_{\mr{P}}$, $\epsilon_{\mr{m}}$ and $\tau$. As long as $N_{\mr{m}}\gg 1$ the problem is however only weakly sensitive to the precise values.

The problem is closed by appropriate boundary conditions at either reservoir. At the left reservoir the potential is set to zero and at the right reservoir the potential is set to $-V_0$. At both reservoirs the normalized concentrations take the values $c_{\pm}=1$, $\cH=\cOH =n$.

\section{Analytical treatment}
\seclab{ana}

In this section we derive analytical expressions for the potential $\phi$ and concentration fields $c_i$ given as functions of the salt and water-ion currents $J_+$ and $\Jw$. As a result of the analysis, we find simple scaling laws for some of the characteristic features in the problem.

\subsection{Basic analysis}

For the simple system without water-ions we know from Refs.~\cite{Smyrl1967, Chu2005, Nikonenko2010} and numerical simulations that the solution in the left compartment is composed of three regions: a locally electroneutral diffusion layer, an extended space-charge region (ESC) and a quasi-equilibrium electric double layer (EDL). Such a solution is sketched in \figref{System2}(a).

Initially, we only consider the left compartment outside the EDL. i.e. the region extending from 0 to $L^*$ in the inset of \figref{System2}(a). In the analysis we will assume that $L^*=L$, which is a good assumption for most parameter values. We introduce the effective water ion density $\cw$ and write
\bsubal
\cw &\equiv \beta \cH +\cOH,\\
2\Jw &= \pp_x \cw - \cw\pp_x \phi - 2\beta \pp_x \cH \approx \pp_x \cw - \cw\pp_x \phi . \eqlab{Jw}
\esubal
We can discard the $2\beta \pp_x \cH$ term because the hydroxide concentration is very much larger than the hydronium concentration in the entire LEN region, and in the ESC region, where this may not be the case, diffusion plays a negligible role compared to electromigration.

We assume that the membrane is completely impenetrable to anions, so that $J_-=0$. The results can readily be generalized to the case of $J_- \neq 0$.

Subtracting \eqref{Jm} from \eqref{Jw} we obtain
\bsubal
2\Jw &=  \pp_x(\cw+c_-)-(\cw+c_-)\pp_x \phi \\
&\approx \pp_x c_+-c_+\pp_x \phi + 2\blamDsqr \pp_x^3\phi-2\blamDsqr \pp_x\phi \pp_x^2\phi, \eqlab{Jw2}
\esubal
where \eqref{Poisson} with $(1+\beta)\cH\approx 0$ has been used. This might introduce an error as we have just argued that $\cw \gg \cH$ does not necessarily hold in the ESC. The majority of the charge density in the ESC does however derive from the salt ions, so reasonable results may still be obtained with this approximation, as verified by our numerical simulations in \secref{numerics}.

 Adding (subtracting) \eqref{Jw2} to (from) \eqref{Jp} we obtain
\bsubal
J_++\Jw &= -c_+\pp_x \phi+\blamDsqr \pp_x^3\phi -\blamDsqr\pp_x\phi \pp_x^2 \phi,  \eqlab{JppJw}\\
J_+-\Jw &= -\pp_x c_+-\blamDsqr \pp_x^3\phi +\blamDsqr\pp_x\phi \pp_x^2 \phi.
\esubal
The second of these equations is easily integrated
\begin{equation}
(J_+-\Jw)x-1 = - c_+-\blamDsqr \pp_x^2\phi +\frac{\blamDsqr}{2}(\pp_x\phi)^2, \eqlab{JpmJwx}
\end{equation}
where the integration constant is set to $-1$ because $-\blamDsqr \pp_x^2\phi +\frac{\blamDsqr}{2}(\pp_x\phi)^2\ll 1$ at $x=0$. The analysis can be carried out without making this simplification, but it makes the resulting expressions less transparent and the effect is only important for very large $\blamD$, e.g., $\blamD \gtrsim 0.1$.

Multiplying \eqref{JpmJwx} by $\pp_x \phi$ and subtracting it from \eqref{JppJw} we obtain a single ordinary differential equation for the potential $\phi$,
\begin{align}
J_++\Jw- \big[(J_+-\Jw)x-1\big] \pp_x \phi=\blamDsqr \pp_x^3\phi-\frac{\blamDsqr}{2}(\pp_x \phi)^3.  \eqlab{Master_eq}
\end{align}
This equation has previously been derived in various forms, for instance in Refs.~\cite{Smyrl1967, Chu2005, Urtenov2007}. A common way of deriving solutions to this equation is to use the method of matched asymptotic expansions \cite{Newman1965,Chu2005, Olesen2010,Yariv2009}. We will use a slightly simpler approach which omits the EDL, while still capturing the essential physics of the problem.

Let us consider the magnitude of the terms in \eqref{Master_eq} in each of the distinct regions. In the electroneutral diffusion layer only the terms on the left of \eqref{Master_eq} matter, since the entire right-hand side stem from the Poisson equation. In the ESC the charge density can obviously not be neglected, and the terms on the right-hand side come into play. The right-hand side terms scale as $\blamDsqr \pp_x^3 \phi \sim \blamDsqr \frac{\Delta \phi}{\Delta x^3}$ and  $\blamDsqr (\pp_x \phi)^3 \sim \blamDsqr \frac{\Delta \phi^3}{\Delta x^3}$, where $\Delta x$ and $\Delta \phi$ is the width of the ESC and the potential drop over the ESC, respectively. Because the conductivity in the ESC is small (few charge carriers), the potential drop over the ESC will be large. It follows that $\blamDsqr (\pp_x \phi)^3 \gg \blamDsqr \pp_x^3 \phi $, and it is therefore reasonable to neglect the $\blamDsqr \pp_x^3 \phi $ term in \eqref{Master_eq}. We then end up with a simple algebraic equation for the electric field, valid in the left compartment outside the EDL
\begin{equation}
1+ \frac{\blamDsqr}{2(J_++\Jw)}(\pp_x \phi)^3 = \left[\frac{J_+-\Jw}{J_++\Jw}x-\frac{1}{J_++\Jw}\right]\pp_x \phi.  \eqlab{algebraic_eq1}
\end{equation}
Since $2\blamDsqr \pp_x^3 \phi = - \pp_x \rho_{\mr{el}}$ the above assumption corresponds to assuming a quasi-uniform distribution of the charge density. This method of simplifying the problem has previously been used by Urtenov \emph{et al.} \cite{Urtenov2007} and dubbed the assumption of quasi-uniform charge density distribution. However, so far this assumption has only been used to simplify numerical calculations, and not to obtain analytical solutions.

To simplify the analysis we introduce a scaled electric field $\Eh$ and a scaled position $\xh$, defined by
 \begin{equation}
 \Eh \equiv - B\pp_x \phi, \quad \text{with}\ B\equiv \left(\frac{\blamDsqr}{2(J_++\Jw)}  \right)^{1/3},
 \end{equation}
and
 \begin{equation}
 \xh \equiv \frac{1}{B}\left[ \frac{J_+-\Jw}{J_++\Jw}x-\frac{1}{J_++\Jw}  \right]. \eqlab{xh_def}
 \end{equation}
This enables us to recast \eqref{algebraic_eq1} as
 \begin{equation}
 -1+\Eh^3=\xh \Eh. \eqlab{algebraic_eq2}
 \end{equation}
Before actually solving this equation we can use it to derive some results characterizing the ESC. The scaled charge density $\rhoh =  \pp_{\xh}\Eh$ is found by implicit differentiation, $3\Eh^2 \pp_{\xh} \Eh =\Eh+\xh\pp_{\xh}\Eh$, which results in
 \begin{equation}
 \pp_{\xh}\Eh=\frac{\Eh}{3\Eh^2-\xh}. \eqlab{y_derivative}
 \end{equation}
Differentiating $\pp_{\xh}\Eh$ again, it is found that the point of maximum charge density is at $\xh=0$ and that
\begin{equation}
\mr{max}(\rhoh)=\rhoh(0) = \frac{1}{3}.
\end{equation}
The simple form of this result is due to \eqref{algebraic_eq2} being trivial for $\xh=0$. The scaled charge density can be related to the unscaled charge density using
\begin{align} \eqlab{Unscaled_charge_density}
\rho_{\mr{el}} &= -2\blamDsqr \pp_x^2 \phi = \frac{2\blamDsqr}{B} \pp_x\xh \pp_{\xh}\Eh \nonumber \\
& =  \left(32\blamDsqr \frac{(J_+-\Jw)^3}{J_++\Jw}\right)^{1/3}  \rhoh.
\end{align}

\begin{figure}[!t]
    \includegraphics[]{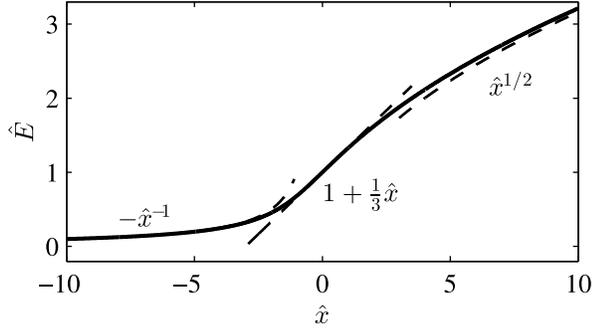}
    \caption{\figlab{y_plot} Plot of scaled electric field $\Eh$ versus scaled position $\xh$ (full line) from \eqref{piecewiseE}. The dashed lines show the limiting behavior for $\xh\rightarrow -\infty$, $\xh \rightarrow 0$, and $\xh\rightarrow \infty$.}
\end{figure}

To proceed beyond this point, we write up the general solution to \eqref{algebraic_eq2}
 \begin{align}
 \Eh= & -\frac{1}{2^{1/3}}\left( -1 +\sqrt{1-\frac{4}{27}\xh^3}\right)^{1/3}e^{i\omega}
 \nonumber \\
 &- \frac{2^{1/3}}{3}\xh \left( -1 +\sqrt{1-\frac{4}{27}\xh^3}\right)^{-1/3}e^{-i\omega},
 \end{align}
where $\omega =0, \frac{2\pi}{3}$ or $\frac{4\pi}{3}$. We require that the solution is real and find
\begin{align}
\Eh = \left\{ \begin{array} {lr} \Eh_- & \text{for } \xh\leq 0 \\
\Eh_+ & \text{for } \xh \geq 0 \end{array} \right. , \eqlab{piecewiseE}
\end{align}
which is continuous and differentiable at $\xh =0$ and where
\begin{align}
\Eh_{\pm} = &\pm\frac{1}{2^{1/3}}\left( \pm 1 \mp \sqrt{1-\frac{4}{27}\xh^3}\right)^{1/3} \nonumber \\
& \pm \frac{2^{1/3}}{3}\xh \left( \pm 1 \mp\sqrt{1-\frac{4}{27}\xh^3}\right)^{-1/3}. \eqlab{ypm}
\end{align}
 In \figref{y_plot} the scaled electric field $\Eh$ is plotted for $-10 < \xh < 10$ along with the asymptotic expressions.

It is noted that although this approach, like the method of matched asymptotic expansions \cite{Bazant2005,Chu2005}, deals with different expressions inside and outside the ESC, the expressions used here are different branches of the same solution and as such they are matched by construction. This is a distinct advantage of the present approach, and it allows for an integration of the electric field to find the potential drop over the system.

We would like to relate the currents to the potential drop rather than the electric field. The task of integrating $\Eh$ is simplified considerably by using \eqref{y_derivative} to make a change of variable
\bsub \eqlab{phi_expressions}
\bal
\phih &=-\int \Eh \ \mr{d}\xh = -\int \Eh \frac{1}{\pp_{\xh}\Eh}\ \mr{d}\Eh =-\int 2\Eh^2+\frac{1}{\Eh} \ \mr{d}\Eh \nonumber \\
&= -\frac{2}{3}[\Eh^3-\Eh^3(\xh_0)]-\ln\left(\frac{\Eh}{\Eh(\xh_0)}\right),
\eal
where $\xh_0\equiv -1/[B(J_++\Jw)]$ (\eqref{xh_def} with $x=0$). Equivalently we define $\xh_1 \equiv (J_+-\Jw-1)/[B(J_++\Jw)]$ (\eqref{xh_def} with $x=1$).

 The unscaled potential $\phi$ is related to the scaled potential $\phih$ as
\begin{equation}
\phi = \int \pp_x \phi \ \mr{d}x = -\frac{1}{B} \frac{1}{\pp_x\xh} \int \Eh\ \mr{d}\xh = j\phih,  \eqlab{phi}
\end{equation}
where $j\equiv (J_++\Jw)/(J_+-\Jw)$ has been introduced for convenience. At the inlet $\Eh$ is small so we can make the approximations $\Eh^3(\xh_0)\approx 0$ and $\Eh(\xh_0)\approx -\frac{1}{\xh_0}$ and find the simpler expression
\begin{equation}
\phi \approx -\frac{2}{3}j\Eh^3 - j\ln\left(-\Eh\xh_0\right). \eqlab{simple_phi}
\end{equation}
\esub
The cation concentration is obtained from \eqref{JpmJwx}
\bsub \eqlab{Concentrations}
\begin{equation}
c_+=\frac{\blamDsqr}{B^2}\left[\frac{1}{2\Eh} +  \frac{1}{j}\pp_{\xh}\Eh \right],
\end{equation}
and since the anions are Boltzmann distributed
\begin{equation}
c_- = e^{\phi} = e^{j\phih}.
\end{equation}
To make the further calculation internally consistent we again use $(1+\beta)\cH\approx 0$, and find from the Poisson equation that
 \begin{equation}
 \cw = c_+-c_--\rho_{\mr{el}} = \frac{\blamDsqr}{B^2}\left[\frac{1}{2\Eh} -\frac{1}{j}\pp_{\xh}\Eh \right]-e^{j\phih}.
 \eqlab{cw_expr}
 \end{equation}
\esub
In conclusion, our model gives analytical expressions for all the relevant fields $\phi$, $c_\pm$ and $\cw$ as function of the position $x$ and the salt and water-ion currents $J_+$ and $\Jw$. This part of the analysis is completely general and does not rely on the specific type of ion-selective interface; the nature of the ion-selective interface is only important for the behavior inside the EDL.

\subsection{The case without water-ion current}

Initially, we consider the simple case of zero water-ion current, $\Jw = 0$. In this limit, the problem only depends on the parameter $\blamD$ and the potential is given by \eqref{phi_expressions}. Since this result gives a closed-form expression for the potential, valid at both under- and overlimiting currents, we consider it to be an extension of earlier asymptotic expressions, valid only in the overlimiting regime, given in Refs.~\cite{Rubinstein1979, Yariv2009}.

To find the approximate dependence on $\blamD$ we consider the limit $\Eh^3(\xh_1)\gg 1$, for which $\Eh(\xh_1)$ is given by $\Eh(\xh_1)\approx \sqrt{\xh_1}$ and the potential at $x=1$ becomes
 \begin{align}  \eqlab{phi_simple_no_Jw}
 \phi(1) &\approx  -\frac{2}{3}\xh_1^{3/2} - \ln(-\sqrt{\xh_1}\xh_0)
 \nonumber \\
 &= -\frac{2}{3} \left( \frac{2(J_+-1)^3}{\blamDsqr J_+^2} \right)^{1/2}
 - \frac{1}{2}\ln\left(\frac{2(J_+-1)}{\blamDsqr J_+^2}\right)
 \nonumber\\
 & \approx  -\frac{2\sqrt{2} }{3} \frac{(J_+-1)^{3/2}}{ \blamD J_+} + \ln\left(\blamD\right).
 \end{align}
The first term on the right-hand side dominates, so for a given overlimiting current the potential drop will roughly scale with $\blamDinv$. This agrees well with the intuitive picture, that the more strictly electroneutrality is enforced, the greater is the potential drop required to create the ESC and drive a current.
In \figref{Analytical_current_no_H} the current is plotted versus the voltage difference for varying $\blamD$. The full analytical solution is shown with a full line and the asymptotic solution is shown with a dashed line (only for $J_+>1$).

\begin{figure}[!t]
    \includegraphics[]{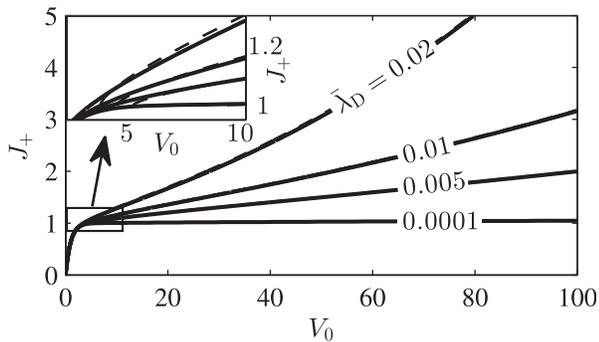}
    \caption{\figlab{Analytical_current_no_H} Salt current $J_+$ plotted versus voltage $V_0$ in the case of vanishing water-ion current $\Jw$. Full line is the analytical expression~(\ref{eq:phi_expressions}), and the dashed line is the asymptotic expression~(\ref{eq:phi_simple_no_Jw}). Only near the limiting current $J_+\gtrsim 1$ do the two cases deviate appreciably (see the insert).}
\end{figure}

\subsection{The influence of water ions} \seclab{water_ions}

To find a relation between $J_+$ and $\Jw$, when water ions are taken into account, we need another constraint on one of the fields. It is however not apparent which constraint we should use or, for that matter, that a simple and physically justified constraint even exists. In the numerical simulations, as we shall later see, the value of $\Jw$ is determined self-consistently by simply requiring continuity of the fields through the membrane. The analytical model does however break down in the EDL, so this method of constraining $\Jw$ cannot be employed here.

Instead, we use a boundary condition which is not entirely rigorous, but does have the appeal of being very simple.
Let us consider \eqref{Jw} in the ESC where diffusion is small compared to electromigration,
\begin{align}
2\Jw \approx - \cw \pp_x \phi.
\end{align}
There is a positive charge density in the ESC so the electric field increases for increasing $x$. Because $\Jw$ is divergence-free this in turn means that $\cw$ must decrease for increasing $x$. However, $\cw$ has a minimum value $\text{min}(\cw) = 2\sqrt{\beta}n$ because of the relation \eqref{Water_LEQ}, so at $x=1$ we must always have $\Jw \geq -\sqrt{\beta}n\pp_x \phi$. For all but the lowest currents (whose contributions are negligible), it seems that this is indeed the constraint which creates the water-ion current. i.e. we determine the water-ion current from
 \begin{align}
 \pp_x \phi|_{x=1} &= - \frac{\Jw}{\sqrt{\beta}n}. \eqlab{Jw_BC}
 \end{align}
By inserting this in \eqref{algebraic_eq1} and solving for $J_+$, we find a relation between $J_+$ and $\Jw$
 \begin{equation} \eqlab{Analytical_Jw}
 J_+ = \Jw\frac{1-\sqrt{\beta}n+\Jw +\frac{\blamDsqr}{2\beta n^2}\Jw^2}{\sqrt{\beta}n+\Jw}.
 \end{equation}
Using this relation together with \eqref{phi_expressions}, the current-voltage characteristic for the system can be evaluated for any set of parameters. We note that this boundary condition is the only place where the equilibrium constant enters in the analysis, so a more general treatment allowing the equilibrium constant to vary can be implemented by an appropriate modification of $n$ in \eqref{Analytical_Jw}.

It is instructive to consider some limiting cases. For overlimiting currents, where $\Jw \gg \sqrt{\beta}n$, \eqref{Analytical_Jw} yields a simple expression for $\Jw$ in terms of $J_+$
\begin{equation} \eqlab{simpler_Jw}
\Jw \approx \frac{\beta n^2}{\blamDsqr}\left(-1 + \sqrt{1+\frac{2\blamDsqr}{\beta n^2}(J_+-1)}\right).
\end{equation}
Expanding this in the two limits $\frac{2\blamDsqr}{\beta n^2}(J_+-1) \sim \frac{\blamDsqr}{n^2} \ll 1$ and $\frac{2\blamDsqr}{\beta n^2}(J_+-1)\sim \frac{\blamDsqr}{n^2} \gg 1$, we find
 \begin{equation} \eqlab{Simple_Jw_limits}
 \Jw \approx \left\{ \begin{array}{lr} J_+-1, & \text{for}\ \frac{\blamDsqr}{n^2} \ll 1, \\[2mm]
 \frac{\sqrt{2\beta}n}{\blamD}\sqrt{J_+-1},  & \text{for}\ \frac{\blamDsqr}{n^2} \gg 1. \end{array} \right.
 \end{equation}

The first of these limits we denote the Kharkats limit, since he studied exactly the situation $J_+ = 1+ \Jw$ where the overlimiting current is only due to screening by water ions \cite{Kharkats1979}. The potential drop over the system is given by \eqref{simple_phi}, and using that $\Eh(\xh_1)=-B\pp_x \phi|_{x=1} = B\frac{\Jw}{\sqrt{\beta}n}$ we find
 \begin{align} %\eqlab{Jw_phi_anal}
 \phi(1) \approx & \frac{2}{3}j\left(-B\frac{\Jw}{\sqrt{\beta}n}\right)^3 - j\ln\left(-B\frac{\Jw}{\sqrt{\beta}n}\xh_0\right)
 \nonumber\\
 =&-\frac{\blamDsqr}{3(J_+-\Jw)}\left(\frac{\Jw}{\sqrt{\beta}n}\right)^3
  - j\ln\left(\frac{\Jw}{\sqrt{\beta}n}\frac{1}{J_++\Jw}\right). \eqlab{phi_vs_Jw}
 \end{align}
An interesting feature of this result is that even in the Kharkats limit $\blamDsqr/n^2 \ll 1$, where the entire overlimiting current is due to water-ion screening, the potential depends on $\blamD$.

\begin{figure}[!t]
    \includegraphics[]{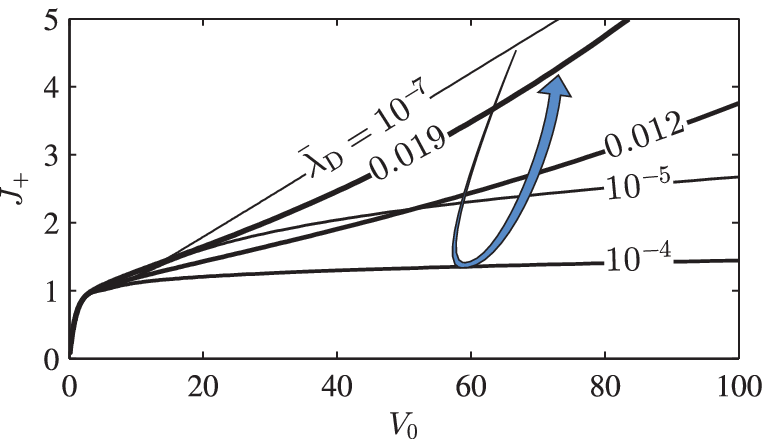}
    \caption{\figlab{Analytical_current_with_H_ink} Salt current $J_+$ from \eqsref{phi_expressions}{Analytical_Jw} plotted versus voltage $V_0$ for $n=10^{-4}$ and $\blamD$ varying from $10^{-7}$ (thin) to 0.019 (thick). The curved arrow indicates the non-monotonous dependence on $\blamD$.  }
\end{figure}

In the case of overlimiting current, the potential drop is determined by inserting \eqref{simpler_Jw} in \eqref{phi_vs_Jw}.
In the Kharkats limit $\blamDsqr/n^2 \ll 1$ given in \eqref{Simple_Jw_limits}, we obtain
 \bsub \eqlab{phi_water_limits}
 \begin{align} \eqlab{phi_high_water}
 \phi(1) &\approx -\frac{\blamDsqr}{3}\left(\frac{J_+-1}{\sqrt{\beta}n}\right)^3
 \nonumber \\
 & \quad -(2J_+-1)\ln\left(\frac{J_+-1}{\sqrt{\beta}n}\frac{1}{2J_+-1}\right),
\end{align}
while in the opposite limit  $\blamDsqr/n^2 \gg 1$ we find
\begin{align} \eqlab{phi_low_water}
\phi(1) \approx  -\frac{2\sqrt{2} }{3} \frac{(J_+-1)^{3/2}}{ \blamD J_+} + \ln\left(\blamD\right).
\end{align}
\esub

A remarkable conclusion can immediately be drawn from these expressions. In the limit $\blamDsqr/n^2 \ll 1$ the potential drop for a given normalized current $J_+$ is seen to increase with $\blamD$. This is opposite to the conclusion in the $\Jw =0$ analysis, and it can be viewed as a result of the coupling between $\Jw$ and $\pp_x \phi$, which is brought about by the boundary condition \eqref{Jw_BC}. We also see that the potential drop scales inversely with $n$ as expected.

In the other limit $\blamDsqr/n^2 \gg 1$, we recover the $\blamDinv$ scaling from the $\Jw=0$ analysis as well as the $\phi$ expression \eqref{phi_simple_no_Jw}. The potential drop over the system will thus have a non-monotonous dependence on $\blamD$. This behavior is seen in \figref{Analytical_current_with_H_ink}, where the salt current $J_+$ is plotted versus voltage for fixed $n$ and varying $\blamD$. It is seen that for some parameter values e.g. $\blamD = 10^{-5}$ we obtain the characteristic S-shaped current-voltage curve found in experiments \cite{Nikonenko2005,Krol1999,Taky1992,Taky1992a}. This indicates, at least on a qualitative level, that the developed model captures the relevant physics of the problem.

%
%In \tabref{Ex_param_values} examples of $\blamD$ and $n$ values are shown for different values of the system parameters $c_0$ and $L$.
%

\subsection{Concentration fields}

The concentration fields found in our analysis exhibit a very rich structure, and it is generally difficult to describe their behavior in simple terms.

\begin{figure}[!t]
    \includegraphics[]{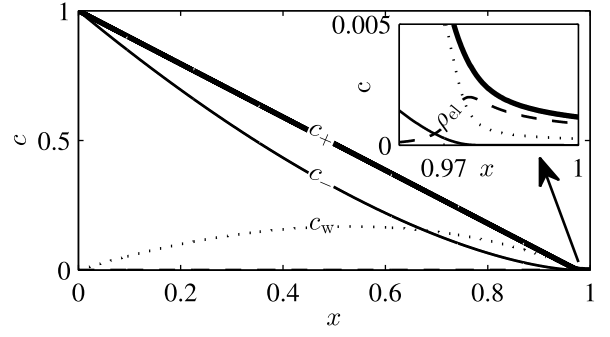}
    \caption{\figlab{Analytical_concentration_fields} The analytical expression \eqref{Concentrations} for the concentration fields $c_+$, $c_-$, $\cw$ and $\rhoe$ plotted versus position $x$ for $n=10^{-4}$, $\blamD=10^{-4}$ and $\Jw = 0.3$. The corresponding salt current is found from \eqref{Analytical_Jw} to be $J_+=1.325$. The insert shows the behavior in the ESC close to the membrane.}
\end{figure}

In \figref{Analytical_concentration_fields} the concentration fields are shown for a given set of parameters. Outside the ESC the fields behave as in the simple LEN theory, with $c_+$ decreasing linearly with $x$, $c_-$ scaling as $(c_+)^j$, and $\cw$ given by the difference $c_+ - c_-$,
 \bsubal
 c_+ &\approx 1-(J_+-\Jw)x,\\
 c_- &\approx \left[1-(J_+-\Jw)x\right]^j,\\
 \cw &\approx 1-(J_+-\Jw)x-\left[1-(J_+-\Jw)x\right]^j.
 \esubal
Since $\Jw$ is determined via \eqref{Analytical_Jw}, these seemingly simple relations do in fact have a quite complicated dependence on all of the parameters $J_+$, $\blamD$ and $n$. It is seen that the expressions break down for $x> 1/(J_+-\Jw)$ and in the following we will let the point $\xrho =1/(J_+-\Jw)$ define the beginning of the ESC.

In the ESC the existence of a non-zero charge density complicates matters further. The charge density has a peak at the beginning of the ESC
 \bsub
 \begin{equation}
 \rhoe\left(\xrho\right) =
 \frac{2}{3}2^{2/3} \bar{\lambda}^{2/3}_\mathrm{D}(J_+-\Jw)\left( J_++\Jw\right)^{-1/3},
 \end{equation}
and in the ESC it decays as
\begin{equation}
\rho_{\mr{el}} \approx 2^{1/2}\blamD(J_+-\Jw)\left[(J_+-\Jw)x-1  \right]^{-1/2}.
\end{equation}
\esub%
In the limit $\blamDsqr/n^2 \gg 1$, where the influence of water ions is negligible, the expressions simplify as
 \bsubal
 \rhoe\left(\xrho\right)  &\approx  \frac{2}{3}2^{2/3} \bar{\lambda}^{2/3}_\mathrm{D} J_+^{2/3}, \\
 \rhoe &\approx 2^{1/2}\blamD J_+\left[J_+x-1  \right]^{-1/2}.
 \esubal
In this case both the peak charge density and the charge density inside the ESC increase with $J_+$.

In the Kharkats limit $\blamDsqr/n^2 \ll 1$, where $J_+ \approx 1+\Jw$ the charge densities simplify as
 \bsubal
 \rhoe\left(\xrho\right) &\approx \frac{2}{3}2^{2/3} \bar{\lambda}^{2/3}_\mathrm{D}\left( 2J_+-1\right)^{-1/3}, \\
 \rhoe &\approx 2^{1/2}\blamD (2J_+-1)\left[(2J_+-1)x-1  \right]^{-1/2}.
 \esubal
Here the peak charge density surprisingly decreases with increasing $J_+$, but inside the ESC the space charge density increases with $J_+$ as before. Also, in this limit the ESC will be very small since $\xrho = 1/(J_+-\Jw)\approx 1$. The reduction in width and magnitude of the ESC will act to suppress EOI in the $\blamDsqr/n^2 \ll 1$ limit. This is similar to the effect of current-induced membrane discharge as described in Ref.~\cite{Andersen2012}. In the literature it has been reported that EOI sets in around $V_0=20$ \cite{Druzgalski2013}. As seen from \figref{Analytical_current_with_H_ink} and the results in \secref{NumRes} water splitting sets in at a lower voltage, which leads us to believe that a suppression of EOI will in fact occur in this limit.

\subsection{Total potential drop}

The developed analytical model gives a general description, valid for any ion-selective interface, of the inlet compartment outside the EDL. To enable comparison with the numerical simulations of a membrane system, a simple model for the potential drop over the remainder of the system is developed.

Inside the membrane there is a very large density $\Nm$ of immobile negative charges. To screen these charges an equally large density of positive ions accumulates. It follows that the conductivity in the membrane is very large, so that the potential drop over the membrane is negligible compared to the other potential drops in the system. While the potential drop inside the membrane can safely be neglected, the potential drops $\Delta \phi_{\mr{m}1}$ and $\Delta \phi_{\mr{m}2}$ over the two membrane interfaces are in general non-negligible. To determine them, we use the assumption of quasi-equilibrium to relate the concentrations just outside the membrane to the concentrations inside the membrane via a Boltzmann factor. Charge neutrality in the membrane then gives
 \begin{align}
 0&=-\Nm+\epsilon_{\mr{P}}(c_+-c_-+\cH-\cOH) \nonumber \\
 &\approx -\Nm+\epsilon_{\mr{P}}
 \big[c_+(1)+\cH(1) \big] \: e^{-\Delta \phi_{\mr{m}1}},
 \end{align}
where we used that the concentration of anions in the membrane is negligible. The same argument applies to both membrane interfaces, so the total potential change across the membrane is
\begin{align}
\Delta \phi_{\mr{m}} = \ln\left( \frac{c_+(1)+\cH(1)}{c_+(2)+\cH(2)} \right).
\end{align}
In the outlet channel, local charge neutrality is an excellent approximation and the water-ion current is totally dominated by hydronium. We therefore have
\begin{align}
c_++\cH =c_-=e^{\phi+V_0},
\end{align}
and it is readily found that
\begin{align}
1+\left(J_++\frac{\Jw}{\beta}\right)(3-x)=e^{\phi+V_0},
\end{align}
since $2 < x < 3$ in the outlet channel and $\phi(3)=-V_0$.
In conclusion the total potential drop across the entire system is
 \begin{align} \eqlab{V0}
 V_0 &= -\big(\phi(1)-\phi(0)\big)-\Delta \phi_{\mr{m}}- \big(\phi(3)-\phi(2)\big)
 \nonumber \\
 &= -\phi(1) -\ln\left(c_+(1)+\cH(1)\right)	+2\ln\left( 1+J_++\frac{\Jw}{\beta} \right),
 \end{align}
where $\phi(1)$ is given in \eqref{phi_expressions}, $c_+$ and $c_{\mr{H}}$ are given in \eqref{Concentrations} and the relation between the currents is given in \eqref{Analytical_Jw}.

\section{Numerical simulations}
\seclab{numerics}

\subsection{Numerical implementation} \seclab{NumImp}
The numerical simulations are carried out in the commercially available finite element software \textsc{COMSOL Multiphysics} ver.~4.3a. Following Gregersen \emph{et al.}~\cite{Gregersen2009}, the equations (\ref{eq:Currents}), (\ref{eq:Salt_NP}), (\ref{eq:Water_NP}), (\ref{eq:Water_LEQ}), (\ref{eq:Poisson}), and (\ref{eq:Membrane_eq}) are rewritten in weak form and implemented in the mathematics module of \textsc{COMSOL} along with the following boundary conditions: $c_{\pm}(0)=1$, $\cH(0)=\cOH(0)=n$, $\phi(0)=0$ and $c_{\pm}(3)=1$, $\cH(3)=\cOH(3)=n$, and $\phi(3)=-V_0$. To improve the numerical stability of the problem we have made a change of variable, so that the logarithm of the concentration fields have been used as dependent variables instead of the concentration fields themselves.

The code has been successfully validated both against known analytical results in various special cases, and by performing careful mesh-convergence analyses as in Refs.~\cite{Gregersen2009}.

Subsequently, the model system has been solved for $c_0$ increasing from $0.1~\SImM$ to $100~\SImM$ in six steps and for $L$ increasing from $1~\SImum$ to $10~\SImm$ in eight steps. Thus a total of 63 configurations have been investigated. For each set of parameters the bias voltage $V_0$ was varied from 0 to 100 in 160 steps (smaller steps at small $V_0$). In total, this resulted in 10080 data points of which 8056 have an overlimiting current $J_+>1$.

\begin{figure}[!t]
    \includegraphics[]{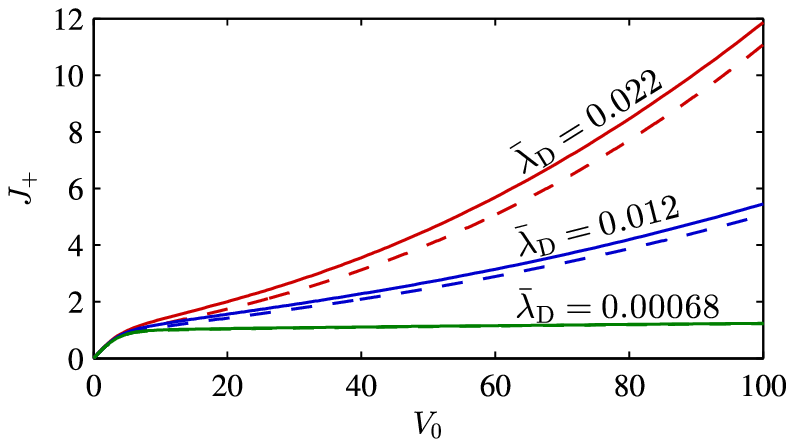}
    \caption{\figlab{Salt_currents_no_H} [Color online] Salt current $J_+$ plotted versus voltage $V_0$ for varying $\blamD$ neglecting the the water-ion current $\Jw$. The full lines are numerical simulations and the dashed lines are the corresponding analytical results from \eqref{V0} with \eqsref{phi_expressions}{Concentrations} inserted.  }
\end{figure}

\subsection{Numerical results} \seclab{NumRes}

Firstly, we present the results for the case without water ions. In this case the problem only depends on one parameter, namely $\blamD$. In \figref{Salt_currents_no_H} the salt current $J_+$ is plotted versus the bias voltage $V_0$ for three values of $\blamD$ (full lines). It should be noted that the normalization current is different for the three cases. The analytical expression from \eqref{V0} with \eqsref{phi_expressions}{Concentrations} inserted is also shown (dashed lines). For small $\blamD$ the current saturates at the limiting current as found in the LEN analysis, while significant deviation from the LEN expression is found for larger $\blamD$ values.
The seen deviations from the LEN expression agrees well with our expectation, that in the limit of very large $\blamD$ a linear $I$-$V$ curve should result.

The analytical $I$-$V$ curves are seen to agree well with the numerical results. The main reason for the small discrepancy is that the width of the EDL becomes non-negligible for large $\blamD$, and therefore the length $L^*$ begins to deviate significantly from the channel length $L$ used in the numerical simulations. There are ways of correcting for this error, but they greatly complicate the expressions. Given that the model already captures the essentials of the behavior we have neglected the inclusion of such corrections at this point.

When water ions are taken into account the problem depends on the normalized equilibrium constant $n = \sqrt{K_{\mr{w}}}/c_0$ and the normalized Debye length $\blamD$. In \figref{currents_low_alpha} the current-voltage curves are plotted for varying $n$ and for two different values of $\blamD$. The analytical expression \eqref{V0} with \eqsthreeref{phi_expressions}{Concentrations}{Analytical_Jw} inserted is also shown with dashed lines.
The light curves shown in the figures are the water-ion currents $\Jw$, and it is seen that the salt currents $J_+$ nearly equal the classical limiting current plus the water-ion current. This is as expected from \eqref{Simple_Jw_limits} since all the considered cases are in the Kharkats limit $\blamDsqr/n^2 \ll 1$. It is seen that several of the curves exhibit the characteristic S-shape found in experiments \cite{Nikonenko2005,Krol1999,Taky1992,Taky1992a}. An interesting observation is that there is a family of curves, an example being the $n=3.2 \times 10^{-4}$ curve in \figref{currents_low_alpha} (a), for which the overlimiting current closely resembles the overlimiting current caused by EOI \cite{Maletzki1992, Rubinstein2002}. These curves are however found in the $\blamDsqr/n^2 \ll 1$ limit where EOI is suppressed. During a measurement series, where the concentration is varied, one might therefore go from a EOI dominated regime to a water-ion current dominated regime, without observing significant qualitative differences in the $I$-$V$ curves.

\begin{figure}[!t]
    \includegraphics[]{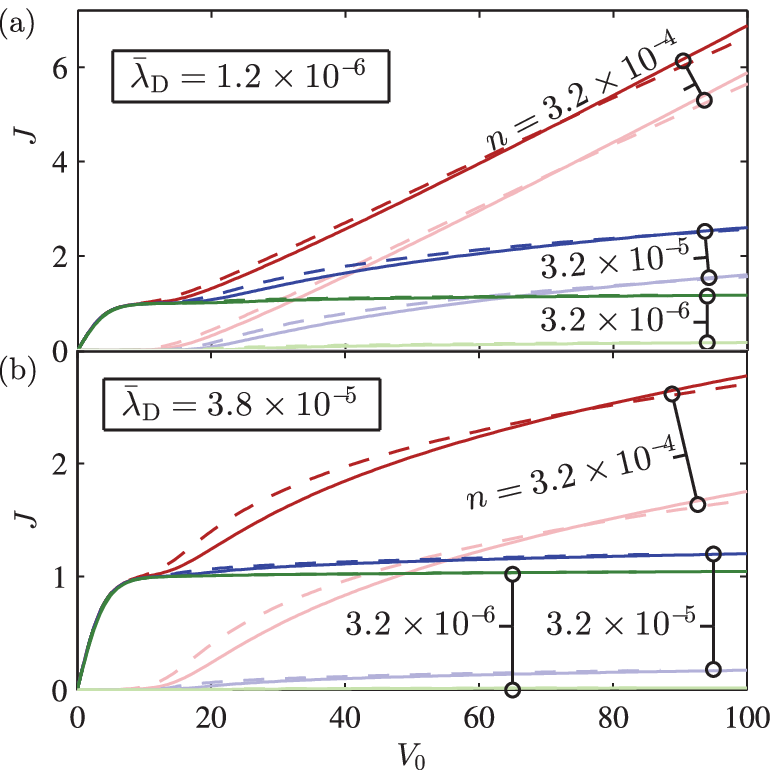}
    \caption{\figlab{currents_low_alpha} [Color online] (a) Salt current $J_+$ (dark) and water-ion current $\Jw$ (light) plotted versus voltage $\VO$ for $\blamD = 1.2\times 10^{-6}$ and $n=3.2\times 10^{-6}$, $3.2\times 10^{-5}$ and $3.2\times 10^{-4}$. The full lines are numerical simulations and the dashed lines are the corresponding analytical results from \eqref{V0} with \eqsthreeref{phi_expressions}{Concentrations}{Analytical_Jw} inserted.  (b) Same as above, but with $\blamD = 3.8\times 10^{-5}$. }
\end{figure}

\begin{figure}[!t]
    \includegraphics[]{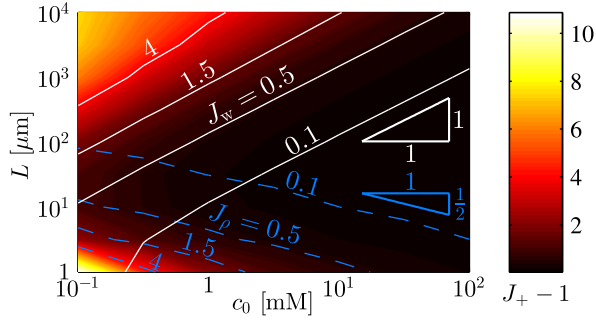}
    \caption{\figlab{Overlimiting_currents_contours} [Color online] Numerically calculated color plot from 0 (black) to 10.8 (white) of the overlimiting salt current $J_+-1=J_{\rho} + \Jw$ at $V_0=100$ as a function of the reservoir concentration $c_0$ and the compartment length $L$. The full lines indicate contours $\Jw=0.1$, $0.5$, $1.5$, and $4.0$. The dashed lines are contours for the current due to the extended space-charge region $J_{\rho}=0.1$, $0.5$, $1.5$, and $4.0$. The slope indications (triangles) show the approximate scalings $L\sim c_0^{-1/2}$ for the $J_{\rho}$ contours and $L\sim c_0$ for the $\Jw$ contours from \eqsref{phi_low_water}{phi_high_water}, respectively. The following parameter values were used in converting from $n$ and $\blamD$ to $c_0$ and $L$: $\epsilon_{\mr{w}} = 6.90\times 10^{-10}\ \SIF/\SIm$, $\VT = 25.8\ \SImV$, $e=1.602\times 10^{-19}\ \SIC$ and $K_{\mr{w}} = 10^{-14}\ \SIM^2$.}
\end{figure}

From the analysis in \secref{water_ions} it is clear that the overlimiting current may be due to either screening by water ions or the development of an extended space-charge region. Which effect is dominant depends on the parameters of the problem. To illustrate this dependence, the overlimiting current at $V_0=100$ is plotted in \figref{Overlimiting_currents_contours} along with contour lines showing the current due to water-ion screening, $\Jw$ (white), and charge neutrality violation, $J_{\rho}=J_+-1-\Jw$ (dark).

In the following we make a more systematic comparison between the analytical model and the results of the numerical simulation. We begin by evaluating the model for water splitting. For each set of parameters $\blamD$, $n$ and $J_+$ used in the simulations the water-ion current $\Jw$ was calculated using \eqref{Analytical_Jw}, and in \figref{Jw_comparison}(a) it is plotted versus the water-ion current which was actually observed in the simulations. Only the cases $J_+>1$ are shown, since $\Jw$ nearly vanishes in the underlimiting regime. It is seen that the developed model captures the majority of the dependence. To better appreciate the level of agreement the simple Kharkats result $J_{\mr{w}}^{\mr{Kha}} =J_+-1$ is shown in \figref{Jw_comparison}(b).

\begin{figure}[!t]
    \includegraphics[]{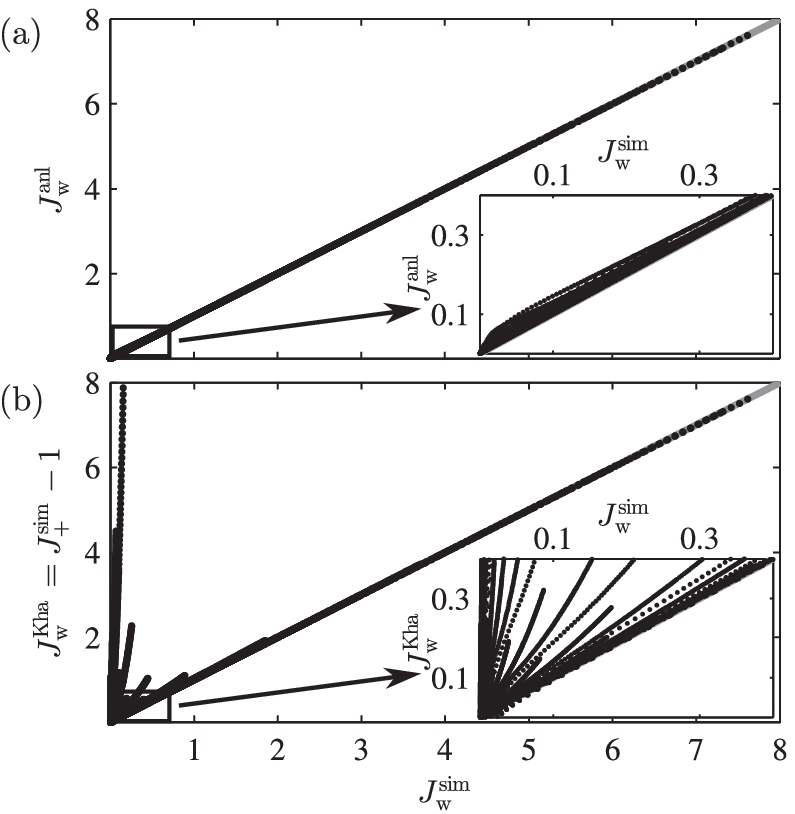}
    \caption{\figlab{Jw_comparison} (a) The analytical water-ion current $J^{\mr{anl}}_{\mr{w}}$ from \eqref{Analytical_Jw} plotted versus the simulated water-ion current $J^{\mr{sim}}_{\mr{w}}$ for the 8056 sets of  values for $\blamD$, $n$ and $J_+$, as defined in the last paragraph of \secref{NumImp}, all having an overlimiting current $J_+>1$. The insert zooms in on the zero-current limit. (b)
Same as above, except that $J^{\mr{anl}}_{\mr{w}}$ is substituted by the Kharkats expression $J_{\mr{w}}^{\mr{Kha}} =J_+^{\mr{sim}}-1$.}
\end{figure}

The total model giving the current-voltage relation for the system has also been evaluated. In \figref{Jcp_comparison} the salt current has been calculated according to \eqsfourref{V0}{phi_expressions}{Concentrations}{Analytical_Jw} and plotted versus the salt current obtained from simulations using the same parameter values. There is seen to be some scatter around perfect agreement between the two models, but the overall behavior is definitely captured by the analytical model.

\begin{figure}[!t]
    \includegraphics[]{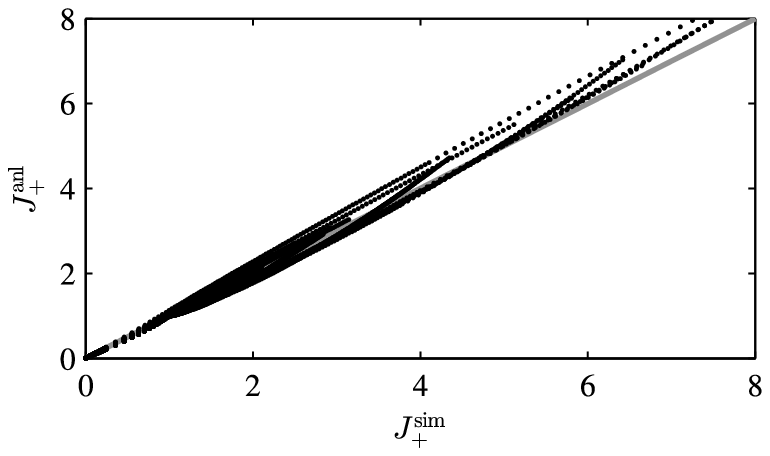}
    \caption{\figlab{Jcp_comparison} The analytical salt current $J^{\mr{anl}}_+$ from \eqsfourref{V0}{phi_expressions}{Concentrations}{Analytical_Jw} plotted versus the simulated salt current $J^{\mr{sim}}_+$ for all 1080 sets of values for $\blamD$, $n$ and $V_0$, as defined in the last paragraph of \secref{NumImp}.  }
\end{figure}

\section{Addition of acid or base}
\seclab{acidbase}

So far we have investigated systems where the ions derive from a dissolved salt. We will now proceed with a more general treatment, where we allow for some concentration of acid $\ca$ or base $\cb$ in the reservoirs in analogy with Ref.~\cite{Kharkats1991}. The acid or base is assumed to be strong so that it dissociates completely, and for simplicity we assume that the conjugate base to the acid is the same as the negative salt ion and that the conjugate acid to the base is the same as the positive salt ion. For instance the salt could be NaCl, the acid HCl and the base NaOH.

Firstly, we consider a system where some concentration $\cb$ of base is added to the system. The ion concentrations are normalized with the total cation concentration at the inlet, i.e. the sum of the salt and the base concentrations. We thus have $c_+(0)=1$, $c_-(0)=1-\cb$ and $\cOH(0)=\cb$. Like in \secref{ana} hydroxide dominates over hydronium, so the relevant transport equation for the water ions is \eqref{Jw}
\begin{equation}
2\Jw  \approx \pp_x \cw - \cw\pp_x \phi,
\end{equation}
but with the difference that $\cw(0) = \cb$ rather than $\cw(0) = (1+\beta)n \approx 0$. We can rewrite the transport equation
 \begin{align}
 2\Jw  &\approx \pp_x \cw - \cw\pp_x \phi
 \nonumber \\
 & = \pp_x(\cw-\cb e^{\phi}) - (\cw-\cb e^{\phi})\pp_x \phi
 \nonumber \\
 &=\pp_x \cw' - \cw'\pp_x \phi,
 \end{align}
where $\cw'\equiv \cw-\cb e^{\phi}$ and $\cw'(0)=0$. The $\cb e^{\phi}$ term behaves exactly like the stationary salt anions, suggesting the introduction of $c_-'\equiv c_-+\cb e^{\phi}$ with $c_-'(0)=1$.

In conclusion the present problem can be mapped onto the problem in \secref{ana}. Adding a base to a system is therefore equivalent to adding a salt of its conjugate acid.
It is noted that to the right of the membrane hydronium dominates the water ion transport, so in this region it will make a slight difference to add a base to the reservoir.

The situation becomes more complex when an acid is added to the reservoir. In this case two quite different situations can result, depending on the amount of added acid. For high acid concentrations the amount of hydronium ions suppress water splitting at the membrane, and the hydronium ions essentially act as a conserved cation. For low acid concentrations hydroxide may begin to dominate the water ion transport at some point and water splitting can occur as in the treatment in \secref{ana}. In \figref{Concentration_fields_with_acid} this situation is illustrated.

\begin{figure}[!t]
    \includegraphics[]{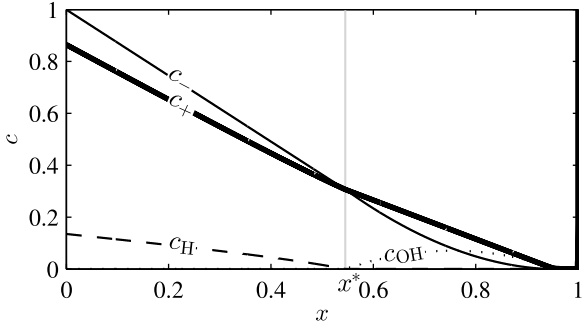}
    \caption{\figlab{Concentration_fields_with_acid} Numerical simulation of the concentrations of salt ions ($c_+$ and $c_-$) and water ions ($c_{\mr{OH}}$ and $c_{\mr{H}}$) plotted versus position $x$ in a system with acid concentration $\ca = 0.135$ and voltage drop $V_0=100$. For $x<x^*$ (left of the vertical gray line), hydronium behaves as a conserved cation, and the system is well-described by a LEN model. For $x>x^*$ hydroxide is the dominant water ion, and the system behaves as the aqueous salt solution analyzed in \secref{ana}. }
\end{figure}

To quantify what is meant by 'high' and 'low' acid concentrations we analyze the system in more detail. From \figref{Concentration_fields_with_acid} it is seen that there are two distinct regions in the solution. To the left hydronium dominates and there is local electroneutrality, while the right part of the channel is equivalent to the system analyzed in \secref{ana}. In the left part of the channel it is easily found that the concentration fields are given as
 \bsubal
 c_- &= e^{\phi} = 1-(J_++\Jw/\beta)x, \eqlab{acid_cm} \\
 \cH &= \frac{\Jw/\beta}{ J_++\Jw/\beta}e^{\phi}+ \left [\ca -\frac{\Jw/\beta}{ J_++\Jw/\beta} \right ]e^{-\phi}, \eqlab{acid_cH}\\
 c_+ &=\frac{J_+}{J_++\Jw/\beta}e^{\phi}+ \left [ 1-\ca-\frac{J_+}{ J_++\Jw/\beta}\right ]e^{-\phi}, \eqlab{acid_cp}
 \esubal
where the concentration fields are normalized with the sum of the acid and salt concentrations at the inlet and $\ca$ is the normalized acid concentration at the inlet. In the limit where there is no water splitting at the membrane the currents are just related via the reservoir concentrations of hydronium and salt cation
\begin{align}
\frac{\Jw/\beta}{ J_+} = \frac{\ca}{1-\ca} \quad, \quad  \text{no water splitting}. \eqlab{no_water_splitting}
\end{align}
If there is water splitting there will be a transition point $x^*$ where the hydronium concentration vanishes. Solving \eqref{acid_cm} and \eqref{acid_cH} for $x^*$ we find
\begin{align}
x^* &= \frac{1}{J_++\Jw/\beta}\left [1 - \sqrt{1-\frac{ J_++\Jw/\beta}{\Jw/\beta}\ca}\right ]. \eqlab{acid_xs}
\end{align}
At that point the salt concentration is
\begin{align}
c^*\equiv c_+(x^*)=c_-(x^*)  =  \sqrt{1-\frac{ J_++\Jw/\beta}{\Jw/\beta}\ca}.\eqlab{acid_cs}
\end{align}
In the right part of the channel the electric field is determined by \eqref{algebraic_eq1} corrected with the new boundary conditions \eqref{acid_xs} and \eqref{acid_cs}
 \begin{equation}
 1+ \frac{\blamDsqr (\pp_x \phi)^3 }{2(J_++\Jw)}= \left[\frac{J_+-\Jw}{J_++\Jw}(x-x^*)-\frac{c^*}{J_++\Jw}\right]\pp_x \phi.
 \end{equation}
Inserting the boundary condition \eqref{Jw_BC} and introducing $G\equiv \frac{\Jw/\beta}{J_+}$ this equation can be recast as a quadratic equation for $J_+$
 \begin{align}
 &\frac{\blamDsqr}{2}\left ( \frac{\sqrt{\beta}G}{n}\right )^3J_+^2-  (1-\beta G)\left ( \frac{\sqrt{\beta}G}{n}\right )J_+
 \nonumber\\
 &= -\left [\frac{1-\beta G}{1+G}\left [1 - c^*\right ]   +c^*\right ]\left ( \frac{\sqrt{\beta}G}{n}\right )+(1+\beta G).
 \end{align}
Just at the point where water splitting begins $G$ will still equal $\frac{\ca}{1-\ca}$ as in \eqref{no_water_splitting} and $c^*$ will be very close to 0. Furthermore, the terms with $\left ( \frac{\sqrt{\beta}G}{n}\right )$ dominate over the term $1+\beta G$, so near that point we can simplify the equation as
\begin{align}
\frac{\blamDsqr}{2}\left ( \frac{\sqrt{\beta}G}{n}\right )^2J_+^2-  (1-\beta G)J_+
\approx -\frac{1-\beta G}{1+G}.
\end{align}
This equation has a solution when the determinant is non-negative, i.e. when
\begin{align}
\frac{\ca}{1-\ca}=G \leq  \frac{1-\beta +\sqrt{(1-\beta)^2 +4\beta(1+2\blamDsqr/n^2)}}{2\beta(1+2\blamDsqr/n^2)}. \eqlab{Critical_ca}
\end{align}
For higher values of $\ca$ there are no solutions which allow for water splitting. The value of $\ca$ for which there is an equal sign in \eqref{Critical_ca}, corresponding to the onset of water-splitting suppression, is denoted the critical acid concentration $\cacrit$. In \figref{Critical_ca} analytical and numerical results for the critical acid concentration are plotted versus $\blamDsqr/n^2$. Numerically the critical concentration is determined as follows. When there is no water splitting the currents are related as in \eqref{no_water_splitting}. The critical concentration is then defined to be the minimum value of $\ca$ for which $\frac{\Jw/\beta}{ J_+} \geq 1.01\frac{\ca}{1-\ca}$, within the voltage sweep interval $0 < V_0 < 100$.

\begin{figure}[!t]
    \includegraphics[]{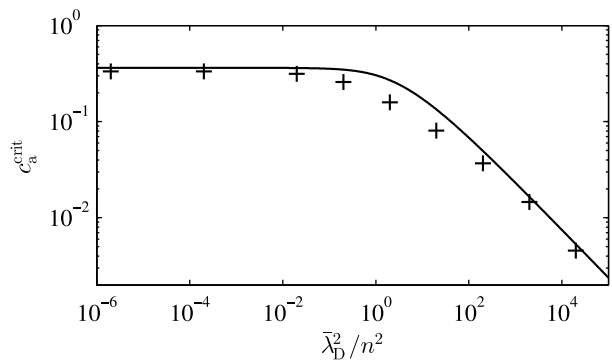}
    \caption{\figlab{Critical_ca} The critical value $\cacrit$ of the acid concentration, corresponding to the onset of water-splitting suppression, plotted versus $\blamDsqr/n^2$. The full line is the analytical expression given in \eqref{Critical_ca}, and the points $'+'$ denote results from numerical simulations.}
\end{figure}

The existence of a critical acid concentration, and its approximate value, is expected to be a robust prediction, which is valid even under circumstances where the assumption of an equilibrated water-dissociation reaction breaks down.

\section{Discussion}
\seclab{discussion}

The results presented in this paper are based on the assumption of a locally equilibrated water-dissociation reaction. Whether this assumption is correct is at present not known, but since our theoretical predictions rely on this assumption, an experimental test of our predictions would constitute a (partial) test of the underlying assumptions.

From the analytical model several useful results are obtained. Our main theoretical result \eqref{phi_expressions} provides the potential $\phi(1)$ at the beginning of the EDL, for a general ion-selective interface with both a water-ion current and the extended space charge region taken into account. In certain limits this result can be simplified to \eqref{phi_water_limits}. The effects of water splitting are accounted for by \eqref{Analytical_Jw}, which provides a relation between the salt current $J_+$ and the water-ion current $\Jw$.

The potential drop across the EDL and the rest of the system depends on the specific ion-selective interface and gives a small correction to the potential. For the specific ion-selective membrane system studied in this work, these corrections are included in \eqref{V0}. The model also provides the detailed structure of the extended space-charge region and yields the simple expression \eqref{Unscaled_charge_density} for the maximum value of the charge density $\rhoe$. The analytical model has been successfully tested against direct numerical simulations, see e.g.~\figref{Jcp_comparison} containing a plot of $J_+^{\mr{anl}}$ versus $J_+^{\mr{sim}}$.

Even if the fundamental assumption of a locally equilibrated water-dissociation reaction is not entirely correct, the analytical model is still useful since it provides an upper bound to the water-ion current, as long as the equilibrium constant $K_{\mr{w}}$ does not change appreciably. For instance, \figref{Overlimiting_currents_contours} shows that in a large portion of the parameter space the influence of water ions is negligible. Since this is an upper bound we can conclude that water-splitting is unimportant for these parameter values regardless of the reaction speed. As described in \secref{water_ions} it would be a relatively simple matter to extend the analysis to allow for a varying $K_{\mr{w}}$.

A strength of the analysis given in this paper is that several of the derived expressions are comparatively easy to test experimentally, since they only depend on a few parameters which can either be estimated or fitted. Consider for instance \eqref{simpler_Jw} for the water-ion current $\Jw$, which in dimension-full terms can be rewritten as $\tilde{J}_{\mr{w}}$,
 \begin{equation} \eqlab{simple_Jw_dim}
\tilde{J}_{\mr{w}} \approx  \frac{2D_{\mr{OH}}}{\gamma D_+}\left(-1 + \sqrt{1+\gamma\left (\frac{\tilde{J}_+}{J_{\mr{lim}}}-1\right )}\right)J_{\mr{lim}},
 \end{equation}
where $J_{\mr{lim}} = 2D_+c_0/L$ is the limiting current, and where $\gamma = (D_{\mr{OH}}/D_{\mr{H}})\: c_0 \epsilon_{\mr{w}} \kB T/(L^2K_{\mr{w}}e^2)$ is a dimensionless parameter. Given knowledge of the reservoir concentration $c_0$ and the length $L$ of the diffusive boundary layer it is possible to calculate $\gamma$ and $J_{\mr{lim}}$ from the definitions. Since \eqref{simple_Jw_dim} is derived under the assumption of an equilibrated water-dissociation reaction, a set of experimental data which fits it, would corroborate that assumption and our model.

Another prediction which can be experimentally tested, is the existence of a critical acid concentration $\cacrit$ for the onset of water-splitting suppression, which may be tested experimentally using the titration method \cite{Zabolotsky1998, Nikonenko2005}. For acid concentrations $\ca$ above $\cacrit$, we predict that the water-ion current and the salt current will be proportional. When $\ca$ is reduced below $\cacrit$, given by \eqref{Critical_ca}, the water-ion current will begin to exceed the value given by \eqref{no_water_splitting}. If, instead, a base is added to the system, we predict that there will be no such critical concentration, and adding an amount of base will in fact be equivalent to adding the same amount of salt. It should be noted that these predictions assume that an added acid or base does not significantly alter the properties of the membrane through chemical reactions. For a chemically stable membrane like nafion this should be a good assumption.

In the analytical treatment it was found that water splitting will act to suppress EOI in the limit of $\blamDsqr/n^2 \ll 1$. We have not verified this prediction by full 3D numerical simulations of EOI, but since water splitting begins at a lower voltage than EOI, it is likely that a suppression of EOI will in fact occur.

Lastly, we emphasize the simplicity and versatility of the employed mathematical method. The reduction of the problem to the simple algebraic equation~(\ref{eq:algebraic_eq2}) for the electric field hugely simplifies the analysis and gives to our knowledge, the first description of the ESC not involving singularities: unlike in the method of matched asymptotic expansions, the fields in this approach do not diverge at the entrance to the ESC, and for this reason closed-form expressions for every relevant quantity can be obtained with ease.

\section{Conclusion}
\seclab{conclusion}

In this paper we have developed analytical and numerical models for the current through and the voltage drop across an ion-selective interface, taking into account both the effect of the extended space-charge region adjoining the interface as well as the effect of water splitting and screening by water ions. Specifically, we have investigated the transport through an ion-selective membrane, but the fundamental results apply to any ion-selective interface.

The fundamental assumption in the analysis is that the auto-dissociation of water happens on a much shorter time scale than the transport of water ions, i.e. we study transport-limited processes. The validity of this assumption is dependent on the particular system under study, but in general the model gives an upper bound to the currents which can be obtained, given a fixed equilibrium constant $K_{\mr{w}}$ for the water-splitting reaction.

In the analytical model the assumption of quasi-uniform charge density distribution has been used to simplify the treatment. The analytical and the numerical model compares favorably and both models exhibit some of the characteristic behavior observed in experiments. The developed analytical model is readily testable in experiments, as it gives both detailed expressions for the current-voltage characteristics, simple scaling laws with few parameters, and predictions about the system behavior upon addition of an acid or a base.

%\bibliography{electrofluidics}

%merlin.mbs apsrev4-1.bst 2010-07-25 4.21a (PWD, AO, DPC) hacked
%Control: key (0)
%Control: author (8) initials jnrlst
%Control: editor formatted (1) identically to author
%Control: production of article title (-1) disabled
%Control: page (0) single
%Control: year (1) truncated
%Control: production of eprint (0) enabled
%

\end{document}